\pdfoutput=1
\documentclass[%
 reprint,
nofootinbib,
 amsmath,amssymb,
 aps,
pra,
floatfix,
]{revtex4-2}

\usepackage[utf8]{inputenc} 
\usepackage[T1]{fontenc}    
\usepackage[table,usenames,dvipsnames]{xcolor}
\usepackage[colorlinks=true,
			hyperindex, 
		    linkcolor = NavyBlue,
		    anchorcolor = hyperrefblue,
		    citecolor = Green,
		    filecolor = hyperrefblue,
		    urlcolor = NavyBlue,
			breaklinks=true]{hyperref}
\usepackage[capitalize]{cleveref}      
\usepackage{url}            
\usepackage{booktabs}       
\usepackage{amsfonts}       
\usepackage{amsmath,mathtools}
\usepackage{nicefrac}       
\usepackage{microtype}      
\usepackage{graphicx}
\usepackage{grffile}
\usepackage{tikz,neuralnetwork}
\usetikzlibrary{positioning}
\usetikzlibrary{shapes,arrows}
\usetikzlibrary{calc,decorations.pathreplacing}

\usepackage[printonlyused,withpage]{acronym}
\makeatletter
\AtBeginDocument{%
  \renewcommand*{\AC@hyperlink}[2]{%
    \begingroup
      \hypersetup{hidelinks}%
      \hyperlink{#1}{#2}%
    \endgroup
  }%
}
\makeatother

\renewcommand{\i}{\ensuremath\mathrm{i}} 
\DeclareMathOperator{\Tr}{Tr} 

\newcommand{\NN}{\mathbb{N}}

\newcommand{\mc}[1]{\mathcal{#1}}

\renewcommand{\vec}[1]{\mathbf{#1}} 

\DeclarePairedDelimiterX{\abs}[1]{\lvert}{\rvert}{%
  \ifblank{#1}{\,\cdot\,}{#1}
}   

\DeclarePairedDelimiterX\norm[1]\lVert\rVert{%
  \ifblank{#1}{\,\cdot\,}{#1}
}   

\DeclarePairedDelimiter{\ket}{\vert}{\rangle}

\DeclarePairedDelimiterX\braket[2]{\langle}{\rangle}%
  {#1\kern0.15ex\delimsize\vert\kern0.15ex\mathopen{}#2}

\DeclarePairedDelimiterX\ketbra[2]{\vert}{\vert}%
  {#1\kern0.15ex\delimsize\rangle\delimsize\langle\kern0.15ex\mathopen{}#2}

\DeclarePairedDelimiterX\sandwich[3]{\langle}{\rangle}%
  {#1\,\delimsize\vert\kern0.15ex\mathopen{}#2\kern0.15ex\delimsize\vert\kern0.15ex\mathopen{}#3}

\DeclarePairedDelimiter\average{\langle}{\rangle}

\newcommand{\myleft}{\mathopen{}\mathclose\bgroup\left}
\newcommand{\myright}{\aftergroup\egroup\right}

\DeclareMathOperator{\Var}{Var} 

\hypersetup{
pdftitle={Scalable approach to many-body localization via quantum data},
pdfsubject={Quantum many-body physics, machine learning},
pdfauthor={Alexander Gresch, Lennart Bittel, Martin Kliesch},
pdfkeywords={MBL, 
						 disorder, Heisenberg model, 
						 adjacent gap ratio, dynamic spin fraction, 
						 entanglement, entropy, 
						 supervised, machine, learning, classification, regression, transfer,
						 scalable, neural networks, recurrent neural networks, RNN, 
						 long short-term memory, LSTM, 
						 gated rectified unit, GRU, 
						 quantum, data, 
						 coefficient of determination
						 },
}

\begin{document}

\title{Scalable approach to many-body localization via quantum data}

\author{Alexander Gresch}
\email{alexander.gresch@hhu.de}

\author{Lennart Bittel}
\author{Martin Kliesch}

\affiliation{%
	Quantum Technology Research Group, Heinrich Heine University D\"usseldorf, Germany
}%

\begin{abstract}

We are interested in how quantum data can allow for practical solutions to otherwise difficult computational problems. 
A notoriously difficult phenomenon from quantum many-body physics is the emergence of \ac{MBL}. 
So far, is has evaded a comprehensive analysis. 
In particular, numerical studies are challenged by the exponential growth of the Hilbert space dimension. 
As many of these studies rely on exact diagonalization of the system's Hamiltonian, only small system sizes are accessible. 

In this work, 
we propose a highly flexible neural network based learning approach that, once given training data, circumvents any computationally expensive step. 
In this way, we can efficiently estimate common indicators of \ac{MBL} such as the adjacent gap ratio or entropic quantities. 
Our estimator can be trained on data from various system sizes at once which grants the ability to extrapolate from smaller to larger ones. 
Moreover, using transfer learning we show that already a two-dimensional feature vector is sufficient to obtain several different indicators at various energy densities at once. 
We hope that our approach can be applied to large-scale quantum experiments to provide new insights into quantum many-body physics. 
 
\end{abstract}

\maketitle

\section{Introduction}
\label{sec:introduction}

The goal of quantum computing is to efficiently solve practically relevant problems that are intractable on classical computers.
Many those problems require a fault-tolerant, universal quantum computer.
This requirement, in turn, comes in conjunction with the need for quantum error correction which yields a daunting overhead in the qubit numbers. 
Both requirements exceed the current available quantum hardware substantially.
Hence, in the meantime, the potential of hybrid quantum algorithms is explored.
They aim to optimally use the few dozens of available qubits with no or little error mitigation schemes.
Most of their pragmatic approaches are centered around \acp{VQA} \cite{Cerezo20VariationalQuantumAlgorithms,Bharti21NoisyIntermediateScale}. 
These algorithms provide heuristics for problems such as finding the ground-state energy in the field of quantum chemistry \cite{Peruzzo2013} or solving combinatorial problems \cite{FarGolGut14}.
Even though the encountered practical constraints impose a tall hurdle,
those efforts appear promising for near-future applications.
Such hopes are furthermore fueled by the achievements in the field of deep learning, especially during the last decade.
Despite the absence of rigorous performance guarantees, there has been a tremendous success of deep learning methods in diverse fields ranging from computer vision, natural language processing to finance and beyond~\cite{JordanMitchell2015review}.
\begin{figure}[b]
\centering 
\tikzset{
  blau/.style = {top color=niceblue!20, bottom color=niceblue!88},
  rot/.style = {top color=red!20, bottom color=red!60}
}

\begin{tikzpicture}[rounded corners=1pt]
    \draw (0.5,0) -- (5.5,0);
    \foreach[count=\j] \start/\length in {-0.4/0.8, 0.6/-1.2, 0.3/-0.6, 1/1, -0.5/1}
    {%
      \coordinate (\j) at (\j,-1);
        \ifnum\j = 4
          \draw[-,gray,line width= 0.15mm] ( $(\j,-0.15) - (0.3,0)$) --++ (0.6,0.3);
          \draw[-,gray,line width= 0.15mm] ( $(\j,-0.15) - (0.2,0)$) --++ (0.6,0.3);
          \node at (\j,-0.8) {\small{$\dots$}};
        \else\ifnum\j = 5
          \fill (\j,0) circle (0.12);
          \draw[-stealth,orange,line width= 0.5mm] (\j,\start) --++ (0,\length);
          \node at (\j,-0.8) {\small{$h_L$}};
        \else
          \fill (\j,0) circle (0.12);
          \draw[-stealth,orange,line width= 0.5mm] (\j,\start) --++ (0,\length);
          \node at (\j,-0.8) {\small{$h_{\j}$}};
        \fi\fi
    } %
    \node at (0.5,-0.8) {\small{$($}};
    \node at (5.5,-0.8) {\small{$)$}};
    \node (h_vec) at (-0.1,-0.8) {\small{$\vec{h}\ \equiv$}};
    %
    \node [draw,rot,
    shape=rectangle,
    alias=rnn_cell, 
    aspect=1,
    scale=4
    ] at (-0.1,-2.5) {};
    \node (rnn_label) at (rnn_cell) {\scriptsize{RNN}};
    \node[below of = rnn_cell,yshift=1mm]{\scriptsize{(feature extractor)}};
    \draw [shorten >=3,shorten <=3,->,thick] (rnn_cell.south) to [out=235,in=205,looseness=3,] (rnn_cell.west);
    
    \draw [shorten >=3,->,thick] (h_vec.south) -- (rnn_cell.north);
    
    \node (h_full) [xshift=1.2cm, right of = rnn_cell] {\footnotesize{Features}};
    \draw [shorten <=3,->,thick] (rnn_cell.east) -- (h_full.west);
    
    \node (feats) [yshift=-0.4cm, gray, above = of h_full] {\footnotesize{Energy density $\epsilon$}};
    \draw [shorten <=3,->,thick,gray] (feats.south) -- (h_full.north);
    
    \node (NN) at (4.3,-2.5) {%
    \begin{neuralnetwork}[layerspacing=9mm,nodespacing=5mm,nodesize=9pt]
        \newcommand{\nodetextclear}[2]{}
        \layer[count=3, nodeclass={input neuron}, bias=false, text=\nodetextclear]
        \layer[count=4, nodeclass={hidden neuron}, bias=false, text=\nodetextclear]
        \linklayers[style={-,thick, shorten <=1pt}]
        \layer[count=3, nodeclass={output neuron}, bias=false, text=\nodetextclear]
        \linklayers[style={-,thick, shorten <=1pt}]
    \end{neuralnetwork}
    };
    \node[below of = NN, yshift=-1mm,xshift=1mm]{\scriptsize{(fully-connected NN)}};
    \draw [decorate,decoration={brace,amplitude=10pt},xshift=-1.8mm,yshift=-0.5cm,thick] (3.3,-2.5) --++ (0,1) node [black,midway,xshift=1cm] {};
    \draw [decorate,decoration={brace,amplitude=10pt,mirror},xshift=0.2cm,yshift=-0.5cm,thick] (5.5,-2.5) --++ (0,1) node [black,midway,xshift=0.1cm] {};
    \node (estimation) at (6.35,-2.5) {$\hat{\vec{y}}$};
    \node (target) [right of = estimation, xshift=-0.5cm] {$|\ \vec{y}$};
    \node [below of = estimation, xshift = 0.33cm,yshift=0.5cm] {\scriptsize{(indicators)}};
    \draw [shorten >=3, shorten <=3, ->, thick] (5.75,0) --++ (1.2,0) -- node[left,rotate=270,xshift=0.7cm,yshift=-0.35cm]{\footnotesize{exact diag.}} ++(0,-2.1);
    
\end{tikzpicture}
\caption{
Workflow for training our model architecture to predict indicator values $\hat{\vec{y}}$ from the system's disorder vector $\vec{h}$.
We pass the latter into a \acl{RNN} as in \cref{fig:rnn_scheme} which extracts general features of $\vec{h}$ in a scalable fashion.
These features can be augmented by the respective energy density $\epsilon$ we are considering.
Together, they are fed into a fully-connected \acl{NN} that maps them to $\hat{\vec{y}}$.
They are compared to the results $\vec{y}$ obtained from exactly diagonalizing the system's Hamiltonian in the corresponding energy density~$\epsilon$.
\label{fig:model_scheme}
}
\end{figure}
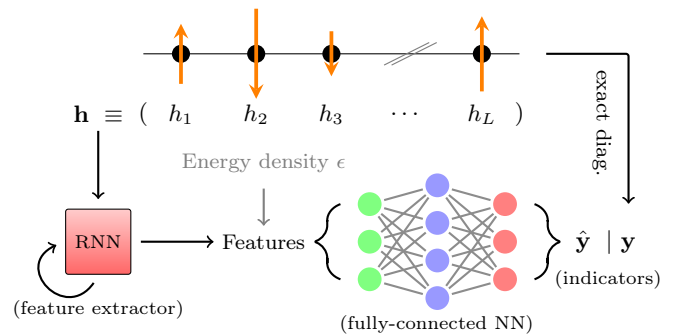%

Over the last year, rigorous performance guarantees for machine-learning-based approaches to quantum many-body physics have been found~\cite{Huang2021a,Huang2021b,Huang2021c}.
These findings suggest that machine learning algorithms are well suitable to generalize efficiently on \emph{quantum data} that is obtained by quantum experiments or a quantum simulation.
In particular, with the recent development in hybrid quantum algorithms such as the \ac{VQE}~\cite{Peruzzo2013,OMalley2015}, variational methods become interesting, viable experimental alternatives.
Alterations to the originally proposed scheme allow for the study of a few eigenvalues and -states around a target energy~\cite{NakanishiSSVQE2018} which does not need to be the ground state~\cite{Higgott2018}.
The \ac{VQE}'s setting suits the study of \ac{MBL} quite well~\cite{Liu2021}.

To demonstrate the importance of the quantum data, difficult problems from quantum physics are needed.
These problems are rendered as such because of their evasive behavior under analytical or numerical analyses.
One of such notoriously difficult problems is the phenomenon of localization in interacting quantum many-body systems, known as \ac{MBL}~\cite{OganHuse2007,PalHuse2010,LuLaflAlet2015}, see e.g.\ Refs.~\cite{NanHus15,EisFriGog15,AletLaf2018} for reviews.
It originates from the well-known Anderson model of non-interacting fermions in a disordered potential where localization occurs above a certain disorder threshold~\cite{Anderson1958}.
The seminal works~\cite{BaskoAleinerAltshuler2006,OganHuse2007} proved the survival of the localization under the introduction of a weak interaction in terms of a perturbation.
This localization can be pinpointed to the emergence of macroscopically many conserved quantities~\cite{Chandran15ConstructingLocalIntegrals,Kim14LocalIntegrals,NanHus15,RadLoukOrt2017,ImbrieRosScar2017}  that suppress the flow of correlations through the system.
In the regime of strong interactions (or conversely, a negligible disordered potential), \ac{MBL} does not occur which indicates a phase transition between the \ac{MBL} phase and the delocalized one.
The latter can be explored deploying e.g.\ classically motivated ergodic 
arguments~\cite{LuiBar2017}.
However, little is known about the transition region between the two phases and its underlying mechanism.
The emergence of \ac{MBL} connects to the fundamental question of thermalization in quantum mechanics~\cite{Deutsch1991,Srednicki1994,Rigol2008}, possibly bridged by the eigenstate thermalization hypothesis (ETH)~\cite{DAlKafPolkRig2015,NanHus15}.
Numerical studies of the transition either apply exact diagonalization~\cite{PalHuse2010} or approximate methods using either shift-invert diagonalization~\cite{PietMaceLuiAlet2018} or renormalization group techniques~\cite{LimSheng2016}.
Around the presumed transition region between the two phases, the numerical methods suffer from the curse of dimensionality because the Hilbert space dimension grows exponentially with the chain length $L$.
Moreover, a numerical extrapolation to the thermodynamic limit at which the transition is expected to be chararacterized by a single value for the critical disorder parameter $h_c$ is hampered by finite-size effects~\cite{KheLimShengHuse2017}.

\subsection{Related works}
\label{subsec:literature}

The idea of applying \acp{NN} to physical problems and, in particular, phase classification, arises as a consequence of its success with feature extraction e.g.\ for conventional image classification, where the classifiers could achieve a higher prediction accuracy than human test groups~\cite{KaimingInit15}.
It has led to a surge of explorations in applying similar methods to difficult problems in (quantum) many-body physics~\cite{NieuwLiHu2016,CarrasquillaMelko2017,LiuNie,Melko2019}.
The phenomenon of \ac{MBL}, in particular, has attracted many numerical approaches using machine learning~\cite{HsuLiDengSarma2018,Venderley2018MachineLearningOut-of-Equilibrium,ZhangWang2019} or deep learning~\cite{SchRegNeu17,NieuwRef2018,HueDauWit18,NieuwBauRef2019}.
The previous attempts typically utilized \acp{NN} for the phase classification in order to extract a phase diagram of the transition in an energy-density- and disorder-parameter-resolved way.
Employing a \ac{RNN} to study the behavior of \ac{MBL} was -- to the best of our knowledge -- first accomplished by Ref.~\cite{NieuwRef2018} who trace the temporal evolution of an observable as a phase classification task.
In variation to those approaches, we propose to employ an \ac{RNN} to characterize a given instance of the Hamiltonian's components in terms of quantum data.
For the characterization, there has been an explorative work done by Nieuwenburg, Baum, and Refael~\cite{NieuwBauRef2019} in the same direction. 
They show the learnability of the adjacent gap ratio by means of convolutional \acp{NN} from the disorder vector joined with the corresponding disorder parameter, i.e.\ from $\Vec{h}\oplus h$~\cite[Appendix]{NieuwBauRef2019}. 
Their efforts, however, resort to a proof-of-principle demonstration and use it for data augmentation.
Moreover, their architecture is not scalable in the system size $L$ because the output size of the convolutional layers grows linearly with $L$.
Such convolutional layers can be made scalable with the input size as demonstrated by Saraceni, Cantori, and Pilati~\cite{Saraceni2020}.
They propose an architecture where the number of extracted features does not grow with the input size and can thus be mapped to a fixed output size.
Apart from this last instance, all the previous methods are restricted to a given, fixed chain length and therefore not applicable to data from a larger system.
Another bottleneck is the fact that the typical input for these approaches consists of heavily preprocessed data such as the entanglement spectrum~\cite{SchRegNeu17} or even a whole eigenvector of the Hamiltonian~\cite{HueDauWit18}.
Both are obtained by exact diagonalization and thus lack a feasible source of training data from the transition regime for system sizes $L\gtrapprox 20$.

\subsection{Our contribution}

In this work, we propose an \ac{NN}-architecture that is both applicable to data from different system sizes and not necessitating any computationally costly preprocessing of the input data. 
We accomplish this by directly presenting the local disorder values $\vec h =(h_1,\dots, h_L)$ to an \ac{RNN}.
This step lifts the system size constraint by treating $\vec h$ as a sequence of inputs such that the sequence length corresponds to the system size.
The output of the \ac{RNN} serves as the extracted feature vector from the disorder sequence.
Typically, such features do not yet resemble the indicators.
Rather, they are global properties of the input which are not tied to a specific regression task.
This view is adapted from results in computer vision where the first layers of image classifying networks merely detect edges and corners, independent of the underlying classification problem~\cite{Goodfellow2016}.
Hence, we use a final fully-connected \ac{NN} as sketched in \cref{fig:model_scheme} that maps the extracted features to the indicator estimates.
With this choice for our architecture, we can investigate in the features further by means of \emph{transfer learning}~\cite{Yosinski2014}.
To this end, we show that a set of features extracted from some indicators can be generalized to other previously unseen indicators.
Moreover, we show that we can achieve this goal with only two features of the input without a significant drop in performance.
Finally, we demonstrate the efficiency of our architecture to enhance the resolution of the phase diagram of the test data set.
We achieve this because our trained network is capable of predicting the indicator values for various choices of the energy density $\epsilon$ and disorder parameter $h$ at once.

We emphasize that this \ac{NN}-based approach to the phenomenon of \ac{MBL} differs from previous attempts drastically.
Previously, \acp{NN} have been used for the classification task of preprocessed inputs~\cite{SchRegNeu17,NieuwRef2018,HueDauWit18,NieuwBauRef2019}.
Such an ansatz depends completely on the availability of the preprocessed input.
We take a step further and demonstrate that distinctive signatures of \ac{MBL}, encoded in the indicator values, are directly learnable from a given disorder realization in a spin chain.
That is, we only enter the defining values of the Hamiltonian and regard the processed indicators as targets, not as inputs to our \ac{NN}.
We obtain these estimates for each disorder realization and for various energy densities at once, i.e.\ we do not require any averages beforehand. 

\subsection{Outline}

In the next Section \ref{sec:neuralnetworks}, we introduce artificial \acp{NN} and in particular our model architecture that is based on a recurrent variant.
We proceed by introducing the quantum many-body system of interest for the study of \ac{MBL} in \cref{sec:model}.
As a test bed for our set-up, this will be the disordered Heisenberg spin chain.
To this end, we present prominent indicators of \ac{MBL} and their behavior in each of the two phases.
In \cref{subsec:approximation}, we demonstrate the scalability of our architecture to predict data for system sizes beyond the training set.
This includes a quantitative benchmark of the quality of the network's output.
As the next step, we emphasize in \cref{subsec:transfer} by the means of transfer learning that the relevant global features of the input are recognized.
Moreover, this hints towards a compatibility between the various indicators which is understood in the study of Anderson localization but remains unclear for \ac{MBL}.
Lastly, we show the numerical efficiency of our method in \cref{subsec:energy_dependency} to obtain a high-resolution phase diagram of the \ac{MBL}-transition.
We complement our work with a summary and an outlook for future directions in \cref{sec:outlook}.

\section{Preliminaries}
\label{sec:preliminaries}

In the following, we start with providing the required background of \acp{RNN}, accompanied by a physical model featuring \ac{MBL}, the Heisenberg spin chain.

\subsection{Recurrent artificial neural networks}
\label{sec:neuralnetworks}

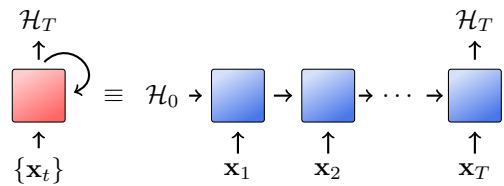
\begin{figure}[b]
{\normalsize
\centering 
\definecolor{niceblue}{rgb}{0.2,0.4,0.9}
\definecolor{NiceGreen}{RGB}{0,153,72}
\tikzset{
  sa/.style = {shading = axis,shading angle=30},
  blau/.style = {sa,top color=niceblue!20, bottom color=niceblue!88, sa},
  gruen/.style = {top color=NiceGreen!10,  bottom color=NiceGreen!70,sa},
  rot/.style = {top color=red!20, bottom color=red!60,sa},
  grau/.style = {top color=black!10, bottom color=gray!80,sa},
  hellgrau/.style = {top color=gray!2, bottom color=gray!20,sa},
}
\begin{tikzpicture}[rounded corners=1pt]

    \node [draw,rot,
    shape=rectangle,
    alias=rnn_cell, 
    scale=3
    ] {};
    
    
    \draw [shorten >=3,shorten <=3,->,thick] (rnn_cell.north) to [out=45,in=23,looseness=3] (rnn_cell.east);
    
    \node (x_full) [below of = rnn_cell] { $\{ \Vec{x}_t \}$};
    \draw [shorten >=3,->,thick] (x_full.north) -- (rnn_cell.south);
    
    \node (h_full) [above of = rnn_cell, scale=1] { $\Vec{\mc{H}}_T$};
    \draw [shorten <=3,->,thick] (rnn_cell.north) -- (h_full.south);
    
    \node (middle) [right of = rnn_cell] {$\equiv$};
    
    \node (h_0) [right of = middle, xshift=-10, scale=1] {$\Vec{\mc{H}}_0$};
    
    \node (h_1) [right of = h_0,draw,shape=rectangle,scale=3,blau] {};
    \draw [shorten >=3,->,thick] (h_0.east) -- (h_1.west);
    
    \node (x_1) [below of = h_1, scale=1] {$\Vec{x}_1$};
    \draw [shorten >=3,->,thick] (x_1.north) -- (h_1.south);
    
    \node (h_2) [right of = h_1,draw,xshift=2mm,shape=rectangle,scale=3,blau] {};
    \draw [shorten >=3,shorten <=3,->,thick] (h_1.east) -- (h_2.west);
    
    \node (x_2) [below of = h_2, scale=1] {$\Vec{x}_2$};
    \draw [shorten >=3,->,thick] (x_2.north) -- (h_2.south);
    
    \node (dots) [right of = h_2,xshift=-0.5mm] {\dots};
    \draw [shorten <=1,->,thick] (h_2.east) -- (dots.west);
    
    \node (h_final) [right of = dots,,xshift=0mm,draw,shape=rectangle,scale=3,blau] {};
    \draw [shorten >=1.5,->,thick] (dots.east) -- (h_final.west);
    
    \node (x_T) [below of = h_final, scale=1] {$\Vec{x}_T$};
    \draw [shorten >=3,->,thick] (x_T.north) -- (h_final.south);
    
    \node (h_T) [above of = h_final, scale=1] {$\Vec{\mc{H}}_T$};
    \draw [shorten <=3,->,thick] (h_final.north) -- (h_T.south);

\end{tikzpicture}
}
\caption{
Scheme of an \ac{RNN} cell as used in \cref{fig:model_scheme}.
On the left, the cell is shown as a black-box that iterates over an input sequence $\{ \Vec{x}_t \}$ and produces an output state $\Vec{\mc{H}}_T$.
Unfolding the cell results in the scheme on the right.
An initial hidden state $\Vec{\mc{H}}_0$ is evolved over $T$ time steps during which the sequence elements are fed into the network one after another.
The final evolved hidden state is released as the network's output.
Each box on the right corresponds to the same cell architecture, i.e.\ having the same weights and biases for each time step.
The recurrent cell can process inputs of arbitrary sequence lengths $T$.
\label{fig:rnn_scheme}
}
\end{figure}%
We use artificial \acp{NN} and in particular their recurrent variant (\acs{RNN}).
\acp{NN} are loosely inspired by their biological counterpart in the human brain.
Effectively, they serve as a black-box approach to a universal function approximator.
They are modularily built by so-called parameterized layers, usually of the form
$\Vec{y}_l = \sigma (W_l \Vec{y}_{l-1} + b_l )$ where the parameters of the $l$-th layer $(W_l,b_l)$ are called weights and biases, respectively.
The linearity is broken by a so-called activation function $\sigma$ which is a non-linear function, usually applied element-wise to its argument.
This way, a predefined type of input $\Vec{x} \eqqcolon \Vec{y}_0$ is processed layer by layer.
This is referred to as the feed-forward pass of the \ac{NN}.
As a consequence, we can consider the \ac{NN} as a parameterized black-box function $f_\theta(\Vec{x}) = \Vec{\hat{y}}$ with parameters $\theta$ given by the weights and biases.
In the \emph{supervised learning} setting, the input $\Vec{x}$ is tied to a target value $\Vec{y}$ of which $\Vec{\hat{y}}$ is an estimation.
The quality of the estimation is quantifiable by the so-called \emph{loss function}.
Its gradient with respect to the network's parameters $\theta$ can be computed efficiently by the method of \emph{backpropagation}.
It is used in an update rule, such as gradient descent, for the parameters to iteratively find a set of parameters that minimizes the loss~\cite{Goodfellow2016}.

The key limitation of the plain-vanilla \ac{NN} is the restriction in the fixed input shape.
\acp{RNN} have a special architecture that allows e.g.\ for an arbitrary input and output length.
This feature is heavily utilized in the field of natural language processing.
The recurrent behavior of a layer is achieved by the introduction of a \emph{hidden state} $\mc{H}$.
To this end, we regard the input $\Vec{x} = (\Vec{x}_1,\Vec{x}_2,\dots,\Vec{x}_T)$ as a sequence of $T$ individual inputs. 
The hidden state can be repeatedly updated according to the network's parameters $\theta$ and the current input, i.e.\ $\Vec{\mc{H}}_t = \Vec{\mc{H}}_t(\theta,\Vec{\mc{H}}_{t-1})$ with $t=1,\dots,T$.
Importantly, the same parameters $\theta$ are used for every update of the hidden state.
The final hidden state $\Vec{\mc{H}}_T$ serves as the output of the recurrent layer.
A schematic is shown in \cref{fig:rnn_scheme}.

\subsection{The model for \texorpdfstring{\ac{MBL}}{MBL}}
\label{sec:model}

A common model often consulted on for the study of MBL is the one-dimensional Heisenberg spin chain of length $L$ whose Hamiltonian reads as
\begin{equation}
	H = J \sum_{i=1}^L \sum_{\alpha\in\{x,y,z\}} \sigma_\alpha^{(i)}\sigma_\alpha^{(i+1)} + \sum_{i=1}^L h_i \sigma_z^{(i)},
	\label{eq:Hamiltonian}
\end{equation}
where $\sigma_{x/y/z}^{(i)}$ denotes the respective Pauli matrix acting on the $i$-th site.
We work with periodic boundary conditions, i.e.\ $\sigma_{x/y/z}^{(L+1)} \equiv \sigma_{x/y/z}^{(1)}$.
The \emph{parameters} $\vec h = (h_1,\dots,h_L)$ are the local disorder strengths which are sampled independently from a uniform distribution over the interval $h_i \in[-h,h]$ for each site $i$.
The variable $h$ is called the \emph{disorder parameter}.
The nearest-neighbor interaction strength $J$ can be set to unity as we are only considering its relation to the value of $h$, i.e.\ we report values for $h$ in units of $J$.

We note that the total magnetization $S_z^\mathrm{tot} \coloneqq \sum_{i=1}^L \sigma_z^{(i)}$ commutes with the Hamiltonian \eqref{eq:Hamiltonian}, and we restrict our considerations to the $S_z^\mathrm{tot}=0$ sector and even chain lengths $L\in2\NN$.
The dimensionality of this sector is $\binom{L}{L/2}$.
This model displays delocalized eigenstates for $h\rightarrow 0$ because the Hamiltonian becomes rotationally invariant in this limit.
On the other hand, i.e.\ for $h\rightarrow\infty$ the interaction term is negligible, and we recover the localization behavior of the Anderson model.
In between these limits, a phase transition from the delocalized phase to the many-body localized one is therefore assumed.
Numerical studies report an estimation of the critical disorder parameter $h_c$ of $h_c \approx 6$\footnote{Due to our definition of \cref{eq:Hamiltonian} via Pauli matrices, the critical value is twice as large as typically reported in the literature.},
which has an additional slight dependence on the considered energy density $\epsilon(E) \coloneqq (E-E_\mathrm{min})/(E_\mathrm{max}-E_\mathrm{min})$~\cite{LuLaflAlet2015}.
This numerically observed so-called \emph{mobility edge} is debated from theoretical grounds and attributed to finite-size effects~\cite{ImbrieRosScar2017}.

There are several properties of the two phases which are shared with the Anderson metal-insulator transition.
Such properties like the system's entanglement or its spectral statistics are typically aimed to be summarized by a single real number.
Since it varies in its numerical value from one phase to the other, it is referred to as an \emph{indicator} for many-body localization.
This is not an order parameter as there exists no mean-field theory for \ac{MBL}~\cite{AletLaf2018}.
Indicators can be divided into three groups of origin: 
(i)~spectral indicators (function of the eigenvalues), 
(ii)~functions of the eigenvectors (e.g.\ entanglement entropies), and 
(iii)~time-averaged observables after a quench.
As one example for a spectral indicator, it is known that the distribution of the spectral gaps of the Hamiltonian varies between the two phases.
In particular, for $h\rightarrow 0$ the gaps are distributed according to the Wigner-Dyson distribution whereas the distribution is Poissonian in the \ac{MBL} phase~\cite{OganHuse2007}.
These two limiting cases are incorporated by the \emph{adjacent gap ratio} $\average{r}$.
This ratio can be computed for the $i$-th spectral gap $\delta_i = E_{i+1}-E_i \geq 0$ as
\begin{equation}
	r_i \coloneqq \frac{\min\{\delta_{i+1},\delta_i\}}{\max\{\delta_{i+1},\delta_i\}}\, .
	\label{eq:adjacent_gap_ratio}
\end{equation}
Averaging over all eigenvalues close to a target energy density and over different disorder realizations yields $\average{r}_{\mathrm{deloc}} \approx 0.53$ in the delocalized limit and $\average{r}_{\mathrm{MBL}} = 2\ln (2) -1 \approx 0.39$ in the \ac{MBL} phase when $h \rightarrow \infty$.

Localization is not only traceable by spectral statistics. 
Another prominent measure is the \emph{half-chain entanglement entropy}~\cite{PalHuse2010}.
To this end, we split the chain in half and calculate the reduced density matrix of the first half $\rho_A := \Tr_B [\rho_{AB}]$ by tracing out the second half of the joint density matrix $\rho_{AB}$.
The density operator is constructed for each eigenstate $\ket{n}$ of the Hamiltonian, i.e.\ $\rho_{AB} = \ketbra{n}{n}$.
The entanglement entropy $\average{S_A}$ is given by computing
\begin{equation}
	S_A \coloneqq \Tr [ \rho_A \ln (\rho_A) ]
	\label{eq:entanglement_entropy}
\end{equation}
and averaging again over eigenstates and disorder realizations.
We normalize this quantity with the expected maximal half-chain entropy which is the Page entropy~\cite{Page1993}.
In this way, the indicator varies from $1$ in the delocalized regime to approaching $0$ in the \ac{MBL} phase as entanglement is suppressed by the local disorder.
Moreover, we note a volume-law scaling of the entanglement entropy with respect to the system size in the delocalized phase but only an area-law scaling in the localized regime~\cite{Eisert2010}.

In addition, the eigenstates carry information about the transport behavior of the spin which is a global conserved quantity.
The \emph{dynamical spin fraction} $\average{\mc{F}}$ quantifies the degree of relaxation of an initial inhomogeneous spin density~\cite{PalHuse2010}.
It is given as
\begin{equation}
	\begin{aligned}
		\mc{F} &\coloneqq 1- \frac{\average{M^\dagger M}}{\average{M^\dagger}\average{M}} \\
		\text{with } M &= \sum_{j=1}^L \sigma_z^{(j)} \exp\left(2\pi\i\ \frac{j-1}{L}\right)
	\end{aligned}
	\label{eq:spin_fraction}
\end{equation}
where the expectation value is taken for all eigenstates close to a target energy.
Again, we average $\mc{F}$ over many disorder realizations.
The persistent spin inhomogeneity in the \ac{MBL} phase means that $\average{\mc{F}} \rightarrow 0$ whereas in the delocalized regime $\average{\mc{F}} \rightarrow 1$.

\section{Results}
\label{sec:results}

In this work, we report on a highly flexible deep learning architecture whose workflow we depict in \cref{fig:model_scheme} that learns the quantum data obtained from an experiment or a numerical study.
In this way, predictions can be made for single instances at various energy levels at once, and we do not need any averages over input configurations.
Moreover, the set-up lifts the restriction of a fixed system size for the available quantum data and only requires the relevant parameters of the underlying Hamiltonian.
We demonstrate that the set-up extracts global, i.e.\ task-independent features from the input which makes it applicable to predicting a broad class of quantum data.
Thus, our approximation scheme serves as a computationally cheap alternative to demanding numerical methods such as exact diagonalization.
We emphasize that, in a broader sense, our method is not limited to the study of \ac{MBL} but applicable to many more problems in quantum many-body physics.

\subsection{Scalable indicator approximation}
\label{subsec:approximation}

Over the last two decades of approaching Anderson localization analytically and subsequently \ac{MBL} mostly numerically, several properties of the phenomenon have been demonstrated to be summarized by the aid of the aforementioned indicators.
We demonstrate that they can be approximated efficiently by an \ac{NN}.
Intuitively, this comes as no surprise for the indicator values are functions of the Hamiltonian's parameters which are taken as the input of the \ac{NN}. 
The defining parameters of the Hamiltonian~\eqref{eq:Hamiltonian} are the local disorder values $\vec{h}=(h_1,\dots, h_L)$ as we consider isotropic nearest-neighbor interactions of relative unit strength.
As we explain later in \cref{subsec:energy_dependency}, our architecture is capable of estimating the indicators for various values of the energy density $\epsilon$ at once.
For now, however, we restrict ourselves to the infinite temperature regime, i.e.\ with $\epsilon = 0.5$ fixed.
In order to accommodate disorder vectors of different lengths, we use an \ac{RNN} architecture that treats the disorder vector as a sequence of the local disorder values.
\acp{RNN} have specifically been designed to handle variable sequence lengths by virtue of their recursive design, see \cref{fig:rnn_scheme} and further details in \cref{appendix:NN_details}.
As loss function, we choose the \ac{MSE} between the obtained estimations of the \ac{RNN} and the actual values obtained by exact diagonalization of the Hamiltonian.
As a framework for setting up the \acp{NN}
and its training, we rely on PyTorch~\cite{paszke2017automatic}.
We publish our data and the code for performing the training of the \acp{NN} and for creating all here presented plots online~\cite{MBLlearning_package}.

\begin{figure}
\centering \includegraphics[width=0.5\textwidth]{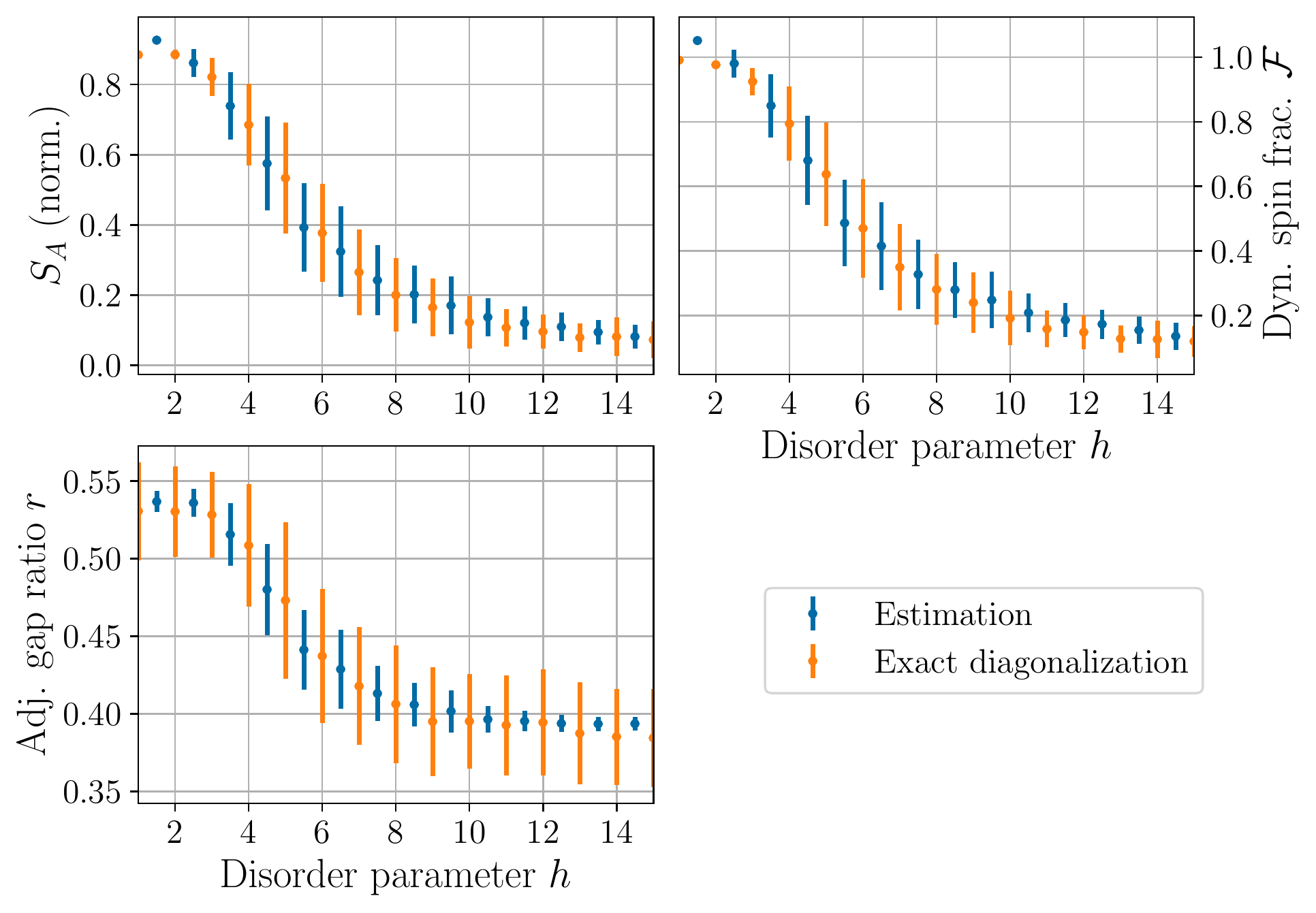}
\caption{
Estimation of the indicator statistics by the \ac{RNN} as a function of the disorder parameter $h$ for the $L=14$ chain at an energy density $\epsilon = 0.5$.
We also provide the respective standard deviations around the means which are reproduced by the \ac{NN} for the first two indicators as well.
\label{fig:fitted_indicators}
}
\end{figure} %
\Cref{fig:fitted_indicators} shows a plot of the learned indicator statistics for $L =14$ where the network has been trained on data from chain lengths $L = 10, 12$. 
We interleave the plotting of the underlying target data with the corresponding output from the \ac{NN}.
For various values of the disorder parameter $h$, we sampled disorder vectors that make up different Hamiltonians.
For each of these, we obtained the vector of indicator values $\vec{y}$ from \cref{sec:model} via exact diagonalization.
Each of the disorder vectors was fed into our \ac{NN} to output an estimation $\hat{\vec{y}}$ of $\vec{y}$.
In the plot, we show the mean and the standard deviation (that results from different realizations of the disorder vector sampled with the same disorder parameter $h$) of $\vec{y}$ and $\hat{\vec{y}}$, respectively.
Especially the entanglement entropy $S_A$~\eqref{eq:entanglement_entropy} and the dynamical spin fraction $\mc{F}$~\eqref{eq:spin_fraction} show a good agreement up to the second moment of the data distribution.
For the adjacent gap ratio $r$~\eqref{eq:adjacent_gap_ratio}, only the mean is well-approximated which indicates that the dependence of $r$ on the level of the particular disorder realization may be harder to learn.
Importantly, we demonstrate that our \ac{NN}-architecture can be queried on data belonging to an arbitrary chain length $L$.
Here, we have trained on smaller system sizes and find a qualitative agreement for the larger system size, $L=14$, in the plot.

Additionally, we can quantitatively benchmark the performance of our network using the \emph{coefficient of determination} $R^2$.
It is used as a benchmarking tool in linear regression 
and is defined as
\begin{equation}
R^2 \coloneqq 1 - \frac{ \sum_i (f(x_i) - y_i)^2 }{\sum_i (y_i - \bar{y})^2} = 1 - \frac{\text{MSE}[f(X),Y]}{\Var [Y]}
\label{eq:coeff_of_det}
\end{equation}
where the sum runs over all data point pairs $\{ (x_i,y_i ) \}$ in the test set, the mean over the targets $y_i$ is denoted by $\bar{y}$, $f$ represents the \ac{NN} and $\Var[Y]$ denotes the variance of $Y$.
So, it essentially compares the \ac{MSE} of the network outputs with the variance in the data.
For a non-linear function $f$
the second term on the right-hand-side is unbounded from above and the corresponding $R^2$ value will lie in the interval $(-\infty,1]$ which is unwanted for a squared expression.
The coefficient of determination \eqref{eq:coeff_of_det} can be transformed to a non-negative number by introducing $R^2_\mathrm{norm.} \coloneqq 1/(2-R^2) \in [0,1] $~\cite{Nossent2012}.
Here, $R^2_\mathrm{norm.} = 1$ means an approximation being exact and $1/2$ constitutes a baseline value, which is attained for $f$ being the constant function that outputs the target mean.
We calculate the normalized coefficient indicator-wise for each value of the disorder parameter $h$.

\begin{figure}[t]
\centering \includegraphics[width=0.5\textwidth]{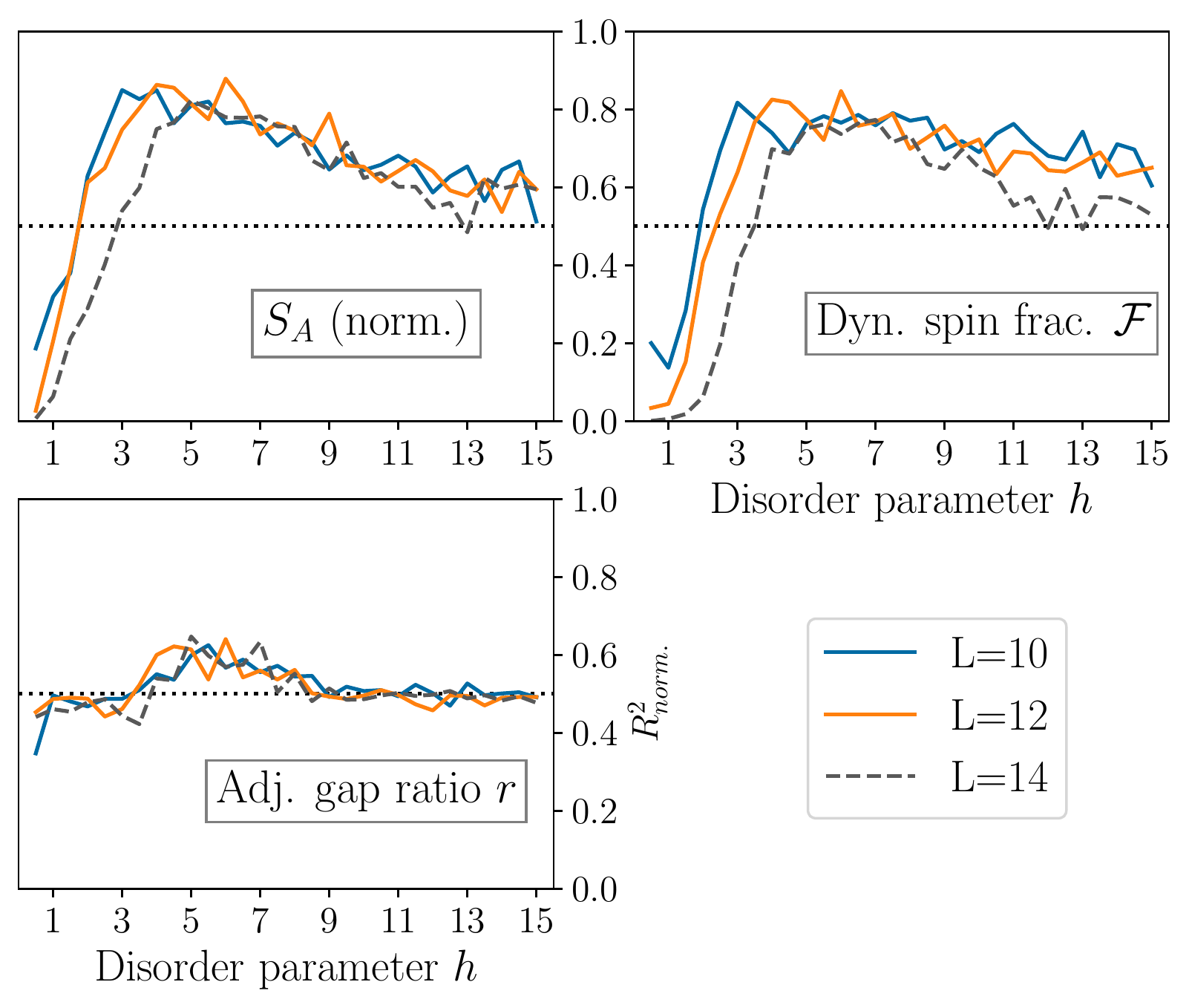}
\caption{
Normalized coefficient of determination $R^2_\mathrm{norm.}$ for each \ac{MBL} indicator as a function of the disorder parameter $h$ at an energy density $\epsilon = 0.5$.
We average all results over five independently trained models.
The network has not encountered any data from the $L=14$ chain (dashed line), yet is capable of capturing the significant part of the indicator statistics.
The dotted line is plotted at $R^2_\mathrm{norm.}=1/2$ to serve as a baseline.
The breakdown of the quality for small disorder parameter values is due to a vanishing variance in the test set which is a consequence of the vanishing disorder in the system, see main text.
\label{fig:coeff_of_det}
}
\end{figure}%

The result for the same energy density as in \cref{fig:fitted_indicators} is presented in \cref{fig:coeff_of_det}.
We emphasize that the network has not encountered any training data from the largest system size, $L=14$.
Yet, it is qualitatively able to estimate values beyond its training set system sizes.
This quantitative observation corroborates our first qualitative one in \cref{fig:fitted_indicators}.
Since the entanglement entropy and the dynamical spin fraction have been well-matched, we see a large value of $R^2_\mathrm{norm.}$ for values $h \gtrapprox 3$ accordingly.
The breakdown for disorder parameter values below that can be attributed to the vanishing variance in the test set for $h \rightarrow 0$ due to the vanishing disorder in the Hamiltonian.
As a consequence, it does not pose a threat to our set-up as it could easily be circumvented by weighting the corresponding training data accordingly.
As we have seen already, the adjacent gap ratio can only estimate the mean of the data distribution faithfully.
Hence, the corresponding normalized coefficient of determination barely exceeds the baseline value.
We attribute this to the unsteadiness in the definition of the adjacent gap ratio caused by the division.
Here, similar Hamiltonians in terms of their respective disorder vectors $\vec h$ can have very different spectra and, in consequence, a very different spectral indicator value.
Moreover, it differs in the limit of vanishing disorder as the spectral indicator can be sufficiently described by the Wigner-Dyson distribution from random matrix theory.
We therefore do not observe a vanishing variance in our numerics which explains the difference in the limit $h \rightarrow 0$ compared to the other indicators.

Lastly, we experimented with the number of required number of samples in the training set.
This is a crucial figure of merit since obtaining the training data always poses a bottleneck in deep-learning approaches to quantum many-body physics.
Since each disorder realization of a given disorder parameter value $h$ is sampled from a uniform distribution over the interval $[-h,h]$, the corresponding variance for a single local disorder strength $h_i$ increases quadratically with $h$.
However, we found no qualitative difference in the approximation quality when considering a training set with a massively increased proportion of data from the \ac{MBL} side.
As the bottleneck of benchmarking our approach is the generation of the training set (due to the cost intensity of the exact diagonalization), we are interested in how the network copes with a shrunken training data set.
We refer to \cref{appendix:further_plots} for the analysis and plots.
In essence, we find that we can shrink the training data set if we allow for more training epochs in return.
This way, we can reduce the training data set down to a number close to the number of trainable parameters in the network.
These observations are crucial for obtaining a data set from an actual experiment in the future where determining indicator values for even a single realization might be expensive.

\subsection{Transfer learning}
\label{subsec:transfer}

The common notion in deep learning is that there exists a hierarchy of abstraction in what the different layers of an \ac{NN} are capable of identifying.
This view has been corroborated by inspecting the first layers of state-of-the-art image classifiers which correspond to edge and corner detection~\cite{Goodfellow2016}.
Since such tasks are detached from the actual classification task, the first layers are said to detect task-unspecific, general \emph{features} of the input and thus regarded as feature extractors.
Only the last layers of a (deep) \ac{NN} map these extracted features to the specific problem at hand.

In this section, we inspect whether such a behavior is exhibited by our proposed model.
We approach this question with the aid of transfer learning~\cite{Yosinski2014}.
The idea is, assuming that the \ac{RNN} actually extracts general features of the disorder vector $\vec{h}$, to keep the \ac{RNN} fixed after we have trained it on a set of \ac{MBL} indicators.
We can now switch the targets in the training set, i.e.\ exchange the target indicators with some new indicators which the network has not encountered before.
As the \ac{RNN}-output is detached from the choice of the target indicators, we only retrain the \ac{NN} that maps the features to the newly chosen indicators.
If the output of the \ac{RNN} corresponds to features of the input that are task-independent, the prediction quality should be comparable to the case where we retrain the full model from scratch on the new data.

\begin{figure}[h]
\centering \includegraphics[width=0.5\textwidth]{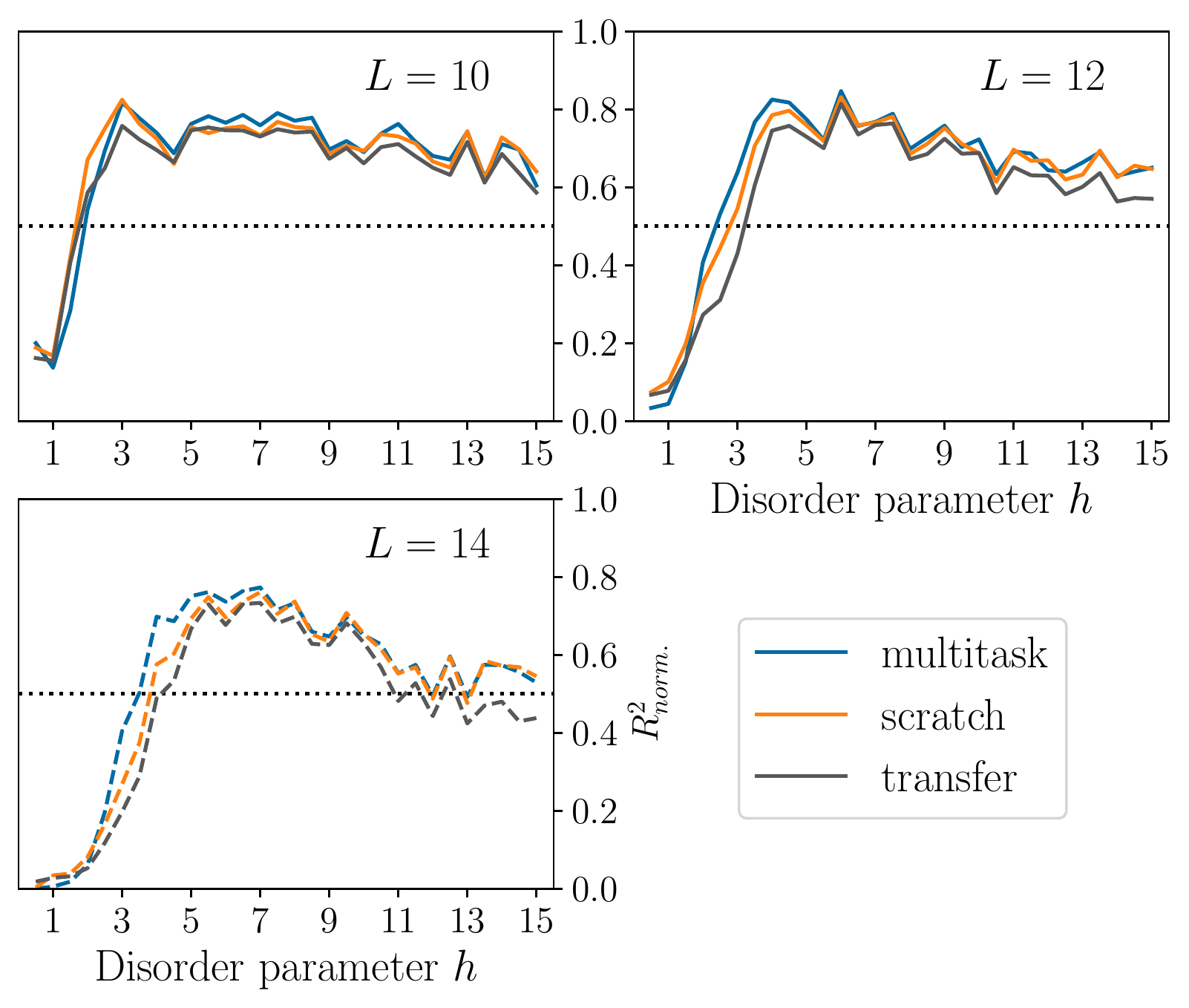}
\caption{
Plot of the normalized coefficient of determination \eqref{eq:coeff_of_det} for the dynamical spin fraction $\mc{F}$.
We compare the model trained via transfer learning (gray line) with an uninitialized model that learns from scratch (orange line) and the previously trained model on all indicators at once (blue line).
Once again, we have excluded data for $L=14$ from the training set which is indicated by the dashed lines in the lower left panel.
The dotted line is plotted at $R^2_\mathrm{norm.}=1/2$ to serve as a baseline.
We averaged the outcome over five independent training procedures.
\label{fig:transfer}
}
\end{figure}

We select the dynamical spin fraction $\mc{F}$~\eqref{eq:spin_fraction} as the transfer target indicator.
To this end, we train our model on the adjacent gap ratio $r$~\eqref{eq:adjacent_gap_ratio} and on the entanglement entropy $S_A$~\eqref{eq:entanglement_entropy} for system sizes $L=10,12$.
Thus, we exclude $\mc{F}$ explicitly from the training set.
Once the training succeeds, we keep the \ac{RNN}'s parameters fixed and only retrain the subsequent \ac{NN} to predict the spin fraction given the output of the \ac{RNN}.
We benchmark the prediction quality with a model of the same architecture that is trained to predict only $\mc{F}$ from scratch.
Furthermore, we compare both predictions with the previous model from \cref{fig:coeff_of_det} that has been trained on all three indicators at once and which we call the multitask network.
A quantitative comparison using the normalized coefficient of determination \eqref{eq:coeff_of_det} is given in \cref{fig:transfer}.
The transferred features lead to a comparable performance as a model that is retrained from scratch and thus tailored to the specific indicator.
Additionally, the performance of the these two networks is very similar to the multitask network.
The differences between any two curves is due to statistical errors.
We find a similar situation when selecting the adjacent gap ratio or the entanglement entropy as the transfer target indicator, respectively (data not shown).
We can attribute the congruence of all three different types of training to the following two reasons.
First, there appears no qualitative difference in the learnability of each of the indicators.
Moreover, they seem to be compatible with each other in the sense that they can all be obtained from the same features.
In our case, we are able to apply the transfer learning scheme using only two features.
We provide more details in \cref{appendix:NN_details}.
This indicates that the extracted features are general enough to allow for the estimation of a variety of indicators which, in turn, do not rely on a specific set of features produced during a specific training procedure.

\subsection{Energy dependency}
\label{subsec:energy_dependency}

\begin{figure}[b]
\centering \includegraphics[width=0.5\textwidth]{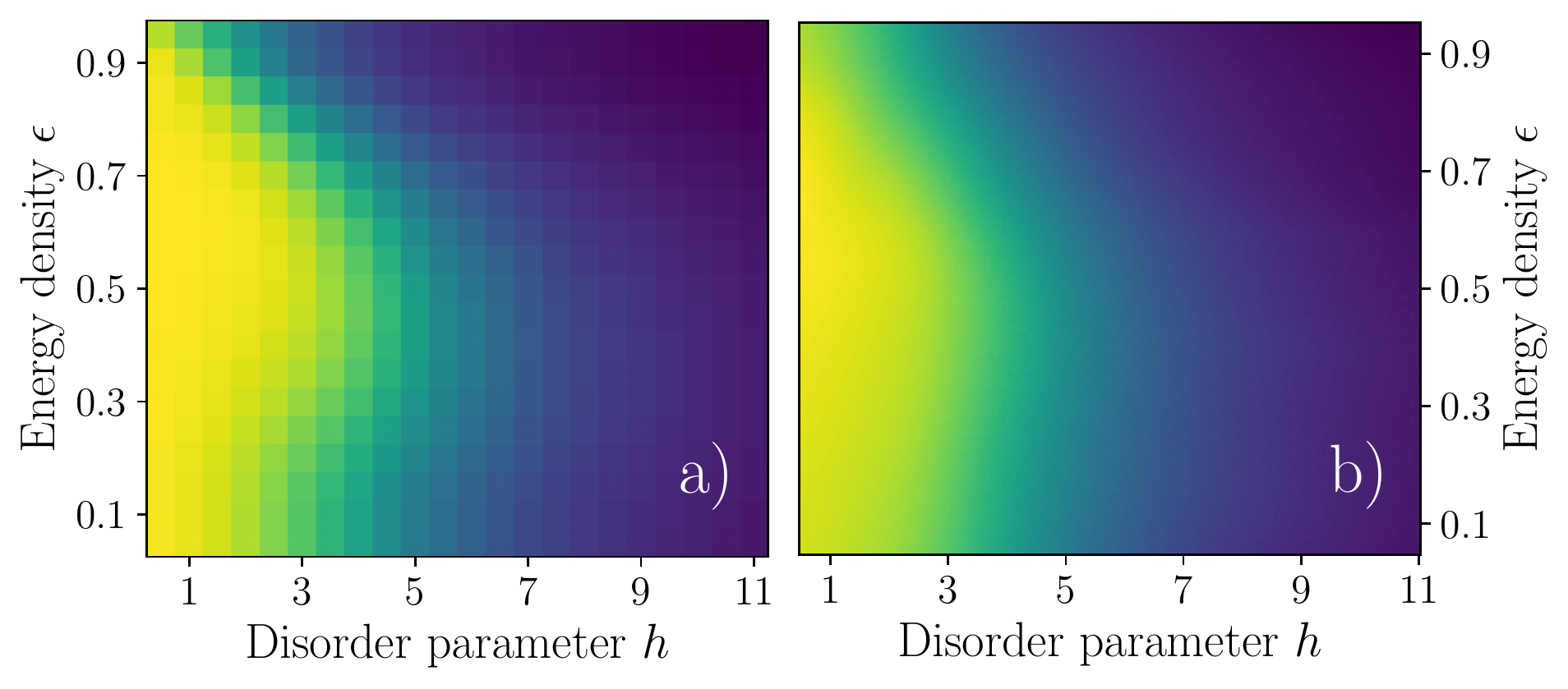}
\caption{
Phase diagram of the dynamical spin fraction $\mc{F}$ for various energy densities $\epsilon$, disorder parameter values $h$ and for a chain length of $L=14$.
In (a), we show the qualitative diagram of the transition obtained by averaging over many disorder realizations.
It is faithfully reproduced by the averaged predictions of the \ac{NN} (b).
Moreover, as the \ac{NN} allows to estimate data for arbitrary values of $\epsilon$ and $h$ we can efficiently increase the resolution of the diagram.
\label{fig:phase_diag}
}
\end{figure}

Lastly, we demonstrate that predictions from our trained estimator recover the results from previous numerical studies of \ac{MBL} in the limit of averaging over many disorder realizations.
Namely, we recover the phase diagram of the transition for various chain lengths $L$ that show the indicator values in dependence of the considered disorder parameter $h$ and energy density $\epsilon$.
To this end, we can generate predictions of unseen trial disorder realizations, i.e.\ random instances of disorder vectors for a given chain length and disorder parameter.
These instances are fed into our \ac{NN} to accumulate a trial data set for various energy densities $\epsilon$ at once.
The latter is straight-forwardly incorporated by augmenting the output of the \ac{RNN} by the corresponding value for $\epsilon$.
Since we solely focus on the network's prediction, we do not need to perform the exact diagonalization procedure for these new instances.
Therefore, generating this large data set is efficient in the system size.
The resulting phase diagram for the dynamical spin fraction $\mc{F}$ is presented in \cref{fig:phase_diag}.

Most importantly, we are now able to generate images of the phase diagram to an arbitrary resolution with numerical efficiacy.
Moreover, we are not limited by the initial resolution in the training data.
This is because we only require forward passes through the \ac{NN} which scales both linearly in the number of queried values for both the disorder parameter and the energy density.
We provide further insights in \cref{appendix:phase_diag}.

%

\section{Conclusion and outlook}
\label{sec:outlook}

We have constructed a \ac{RNN} architecture that approximates values for certain indicators for \ac{MBL} directly from the variable part of the Hamiltonian, i.e.\ the local disorder strengths.
The recurrent set-up ensures that the network can process data for an arbitrary system size $L$ and produce a good estimation output provided the trial system size is not too far off the training set.
Moreover, our approach does not require any further computationally expensive preprocessing of the input data.
In this way, we are able to characterize single disorder realizations by providing the corresponding indicator values.
By inspecting the intermediate features of the \ac{RNN} by means of transfer learning, we observe that all considered indicators can be derived from two features alone.
Furthermore, they serve as an archetype for various indicators at arbitrary energy densities at once.
This enables us to study the transition region by means of phase diagrams that can be rendered to an arbitrary resolution.

\subsection*{Outlook}

With a training set that consists of indicator sets from different system sizes, we envision an interplay between an actual experiment and our architecture.
The experiment can address systems consisting of dozens of spins or qubits.
Thus, it delivers the training set for the architecture beyond what is reachable by exact diagonalization studies.
As we demonstrated, our architecture is not inclined to a specific data type.
Thus, the experiment is not restricted to a certain indicator but can provide the most amenable one (such as the growth of the entanglement entropy~\cite{XuChenFan2018} or the imbalance after a quench~\cite{Schreiber2015,Schreiber2016,XuChenFan2018,KohlBlochAidels2019}) for the training set.
Motivated by our findings in \cref{subsec:approximation}, we conjecture that only a few realizations per disorder parameter are sufficient as to merely guide the extrapolation.
In addition, the indicators are expected to become more and more pronounced in their respective shape.
Therefore, we do not expect large deviations from the case of smaller system sizes up to finite-size effects.
The whole premise of transfer learning relies on the assumption that the additional data for a larger system size only serves as a guidance for the overall learned structure on the training set.
This boosts training the \ac{NN} significantly~\cite{Yosinski2014}.
Given experimental training input, the network can in turn provide estimates for data outside of or in between gaps in the training set which can be benchmarked by the experiment in return \cite{Mohseni21DeepRNNs,Miles2021}.
Other possibilities of enriching the training set is to resort to numerical approximations, for example by tensor networks methods which are well-suited deep within the \ac{MBL} phase~\cite{FriWerBro15} or yet another \ac{NN} architecture to even speed up those methods~\cite{Guo2018}. 
With the data at hand, a more detailed examination of the compability of different indicators allows to shed some light on their yet unknown coaction towards \ac{MBL}.
Diving deeper into the interpretation of the archetypical feature and the compatibility of various indicators is an interesting research direction for future works.

Our proposed scheme aims to bring together the often independent advances in experiments and numerics, and we see possible research directions in the now scalable phase classification task and a better understanding of the learning process of the recurrent feature extractor. 
Furthermore, the connection of our method with a \ac{VQA} is of broader interest ranging from applications in condensed matter and statistical physics to the field of (hybrid) quantum computation or quantum machine learning.
Compared to the existing traditional numerical methods, the interplay of a quantum experiment or its simulation with our method may constitute a new type of quantum advantage in the sense that we can obtain an efficient classical method only via accessing a quantum data set.
Such a pairing provides a potentially powerful computational tool that is yet to be augmented with experimental data in the future.

\begin{acknowledgments}
We thank Christian Gogolin for fruitful discussions. 
Computational support and infrastructure was provided by the ``Centre for Information and Media Technology'' (ZIM) at the University of D\"usseldorf (Germany).
This work has been funded by the Deutsche Forschungsgemeinschaft (DFG, German Research Foundation) under the grant number 441423094 within the Emmy Noether Program. 
\end{acknowledgments}

\newpage
\appendix
\section*{Appendix}
\renewcommand{\thesubsection}{\Alph{subsection}}

In this appendix, we provide more details on our network architecture and the training procedure.
Starting with \cref{appendix:NN_details}, we describe the generation of the data sets and detail the architecture of our approach.
In \cref{appendix:further_plots}, we examine the network's performance under a shrinking data set size.
Finally, we give some more comments on the obtained phase diagram in \cref{subsec:energy_dependency} and its analysis in \cref{appendix:phase_diag}.

\subsection{Details on the network architecture and the training procedure}
\label{appendix:NN_details}

We briefly describe how we set up the training and the test set as well as the network architecture used for the results in the main text.
We set up a grid for the disorder parameter $h$, i.e.\ we chose 30 values $h=0.5,1,1.5,\dots,15$ which lie well around the assumed critical disorder parameter value of $h_c \approx 6$.
For each chain length $L=10,12,14$ we have sampled disorder vectors $\vec{h}$ with entries $h_i$ independently and identically distributed from the uniform distribution, such that $h_i \in [-h,h]$ for a given disorder parameter value $h$. 
For each $h$ and $L$, this was done $N_\mathrm{train}=1000$ and $N_\mathrm{test}=100$ times for the two data sets, respectively.
Each of these disorder vectors yields a realization of the Hamiltonian \eqref{eq:Hamiltonian}.
Its eigenvalues and -vectors were found via exact diagonalization.
We have chosen a grid of $N_\epsilon = 19$ energy densities $\epsilon = 0.05, 0.1, 0.15, \dots, 0.95$ and have kept the $100$ next closest eigenvalues and their corresponding eigenvectors for calculating the three indicators from \cref{sec:model}. 

\begin{figure}[b]
\centering 
\tikzset{
  blau/.style = {top color=niceblue!20, bottom color=niceblue!88},
  rot/.style = {top color=red!20, bottom color=red!60}
}

\resizebox{0.49\textwidth}{!}{%
\begin{tikzpicture}[rounded corners=1pt]
    \node (NN) {%
    \begin{neuralnetwork}[layerspacing=10mm,nodespacing=5mm,nodesize=10pt,toprow=true]
    \tikzstyle{bias neuron}=[neuron, fill=gray!15,alias=bias]
        \newcommand{\nodetextclear}[2]{}
        \newcommand{\nodetexteps}[2]{\ifnum0=#2 $\epsilon$ \else \fi}
        \layer[count=2, nodeclass={input neuron}, bias=true, biaspos=top row, text=\nodetexteps]
        \layer[count=5, nodeclass={hidden neuron}, bias=false, text=\nodetextclear]

		\foreach \i in {1,...,5}{
    \link[from layer=0, to layer=1, from node = 0, to node = \i,style={-,thick, shorten <=1pt,gray!40}]
}
        \foreach \i in {1,...,5}{
    \link[from layer=0, to layer=1, from node = 1, to node = \i,style={-,thick, shorten <=1pt}]
}
		\foreach \i in {1,...,5}{
    \link[from layer=0, to layer=1, from node = 2, to node = \i,style={-,thick, shorten <=1pt}]
}
        
        \layer[count=3, nodeclass={output neuron}, bias=false, text=\nodetextclear]
        \linklayers[style={-,thick, shorten <=1pt}]
    \end{neuralnetwork}
    };
    \node (NN_hidden) [below of = NN, yshift=-5mm,xshift=0mm]{\footnotesize{\texttt{hidden\_size} = 10}};
    \node[below of = NN_hidden, yshift=7mm,xshift=0mm]{\color{gray}{\footnotesize{\texttt{hidden\_size} = 31}}};
    \node[right = of NN, yshift=-1mm,xshift=-10mm]{\footnotesize{\texttt{output\_size} = 3}};
    \node[above = of NN, yshift=-10mm,xshift=-10mm]{\color{gray}{\footnotesize{energy density}}};

    \node [draw,rot,
    shape=rectangle,
    alias=rnn_cell,
    left = of NN,
    aspect=1,
    scale=4
    ] {};
    \node (rnn_label) at (rnn_cell) {\scriptsize{RNN}};
    \node (rnn_in) [below of = rnn_cell,yshift=-3mm]{\footnotesize{\texttt{input\_size} = 2}};
    \draw [shorten <=3, shorten >=3,->,thick] (rnn_in.north) -- (rnn_cell.south);
    \node[above of = rnn_cell,xshift=-4mm]{\footnotesize{\texttt{hidden\_size} = 2}};
    \draw [shorten >=3,shorten <=3,->,thick] (rnn_cell.north) to [out=135,in=145,looseness=3,] (rnn_cell.west);
    \draw [shorten <=3,->,thick] (rnn_cell.east) -- (NN.west);
    
    \node (table) [shape=rectangle,above = of NN,xshift=2.4cm,yshift=-1.5cm,scale=0.85] {%
    	\begin{tabular}{cc}
      & \# parameters \\ \hline
RNN   & 36   \phantom{(220)} \\
NN    & 63   \color{gray}{(220)}\\
Total & 99   \color{gray}{(256)}         
\end{tabular}
};
\end{tikzpicture}
}
\caption{
Details of our model architecture of \cref{fig:model_scheme}.
The \acl{RNN} as in \cref{fig:rnn_scheme} takes in the preprocessed input iteratively.
Afterwards, the final hidden state is fed into the fully-connected \ac{NN}.
It can be augmented by the respective energy density $\epsilon$ as done for \cref{subsec:energy_dependency}.
The corresponding alterations in the network architecture are emphasized by the gray font.
The total number of trainable parameters (including biases) are given in the table.
\label{fig:network_architecture}
}
\end{figure}
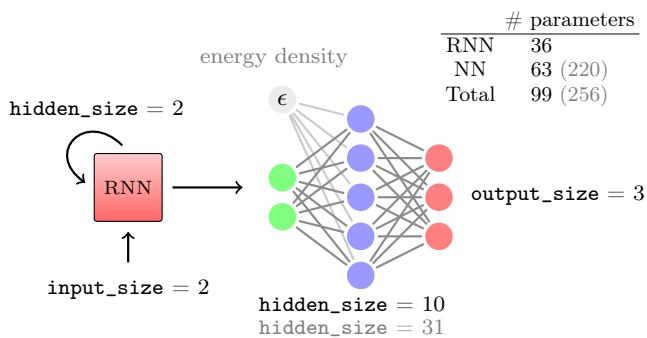%

The architecture of our proposed network scheme is summarized in \cref{fig:network_architecture} and we explain its choice in the following.
The first part consists of an \ac{RNN}-cell that serves as a feature extractor of the input.
The \ac{RNN} is presented each disorder parameter $h_i$ successively and updates its hidden state according to its parameters and the value of $h_i$.
The hidden state was initialized as zero.
After having fed in $h_L$, the final updated hidden state is released as the output of the \ac{RNN}.
We treat this output as the feature vector of the disorder vector.
Due to this recursive procedure, \acp{RNN} can be unstable during training because of exploding or vanishing gradients in the optimization procedure.
In order to circumvent this problem, the \ac{LSTM} cell~\cite{LSTMbook1997} and the \ac{GRU}~\cite{cho-etal-2014-properties} have been proposed with competing performance-efficiency trade-offs~\cite{LSTMGRUcomparison2014}.
We find the latter to be slightly better in performance during training.
Concerning the number of output features of the \ac{RNN}, we find qualitative good results when choosing a feature dimension of $2$.
A larger dimensionality does increase the performance of the indicator approximation, however, we observe a severly decreased performance when applying the transfer learning scheme from \cref{fig:transfer}.
We have only used a single RNN cell of depth one.
Lastly, we have performed a computationally inexpensive preprocessing of the disorder vector.
We regroup the elements of the disorder vector in pairs of two, i.e.\ transform according to $[h_1,h_2,h_3,\dots,h_L] \mapsto [(h_1,h_2), (h_2,h_3), \dots, (h_{L-1},h_L), (h_L,h_1) ]$.
Regroupings into even larger tuples are also possible.
The pairing in two, however, fits in well with the nearest-neighbor interactions and the periodic boundary condition and, furthermore, leads to the best performance.
Afterwards, the feature vector is augmented by the value for the energy density $\epsilon$ under consideration.
Together, we map them to the three indicator values by a fully-connected \ac{NN} of hidden size $10$.
As the loss we choose the \acf{MSE} and train the model for $N_\mathrm{epochs} = 15$ on the training data.
We use the Adam optimizer with default values~\cite{AdamOptim2014}, a batch-size of $128$ and a learning rate $\eta=10^{-3}$.

\begin{figure}[t]
\centering \includegraphics[width=0.5\textwidth]{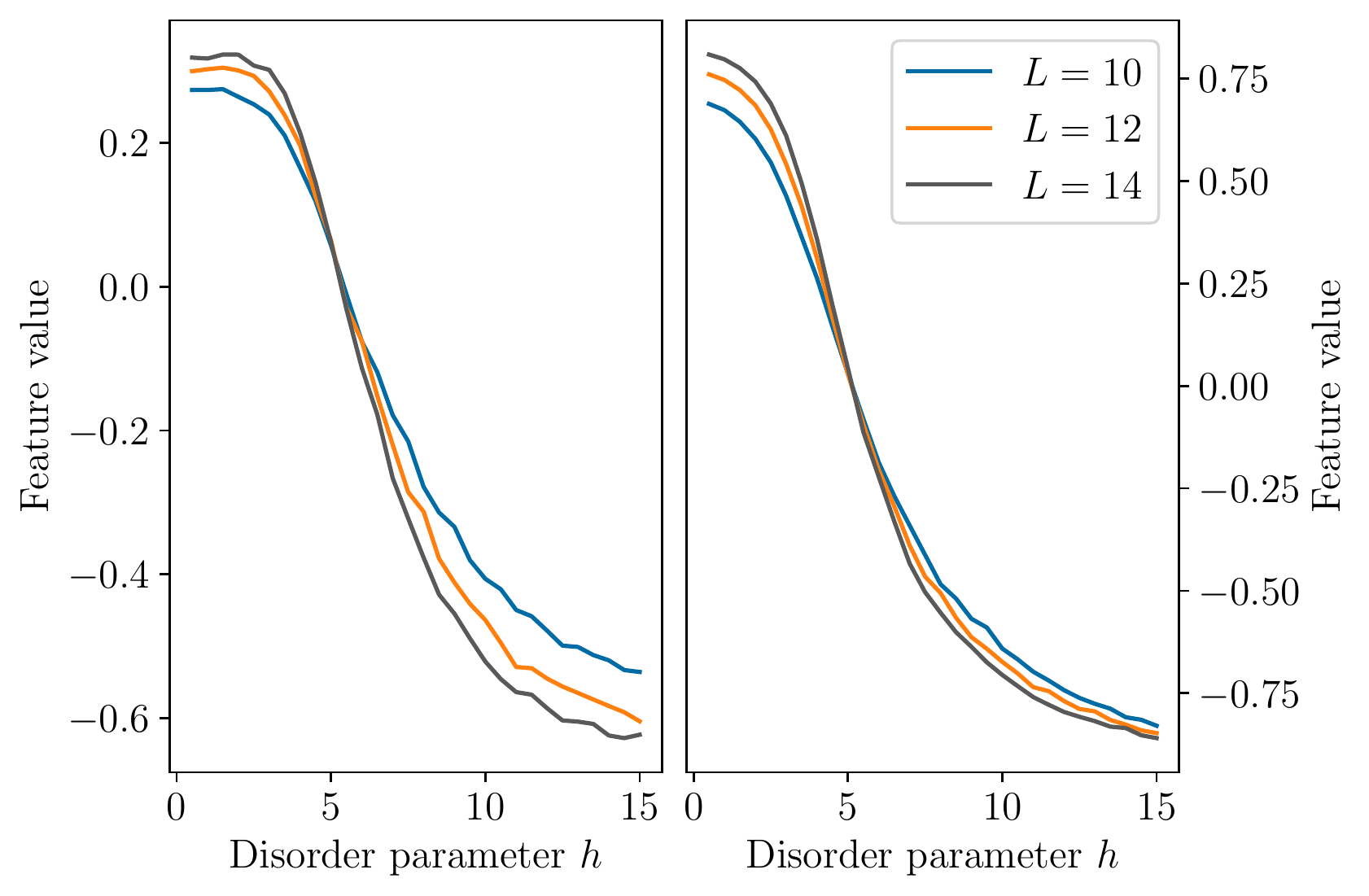}
\caption{
Typical features produced after the training averaged over the training set.
Error on the means are on the scale of the line thickness.
The crossing points vary from training to training which makes retraining and averaging a necessity.
\label{fig:typical_feature}
}
\end{figure}

For the transfer learning scheme of \cref{subsec:transfer} and for creating the model that is capable of dealing with an arbitrary energy density $\epsilon$ in \cref{subsec:energy_dependency}, the training consists of two stages: we first proceed as outlined above
This pretraining is necessary to facilitate an easier focussed training of the \ac{RNN} to extract meaningful features which we show in \cref{fig:typical_feature}.
Then, we fix the parameters of the \ac{RNN} and thus the intermediate features, and train the subsequent fully-connected \ac{NN} on the full training data for $30$ more epochs with a decreased learning rate of $10^{-4}$ following the Adam optimizer routine.
This fine-tuning of the \ac{NN} yields a greater performance compared to training the two components of the model jointly.
The choice for the hyperparameters (architecture of the two individual components, feature size, number of hidden neurons and the optimizer parameters) above has been determined on a held-out validation data set.

\subsection{Examination of the data set size}
\label{appendix:further_plots}

In this section, we provide details on the results of \cref{subsec:approximation}.
In particular, we investigate the performance dependence on the size of the training data set.
We can test this quantitatively by decreasing the number of samples per disorder parameter $N_\mathrm{train}$.
In this setting, half a value in $N_\mathrm{train}$ corresponds to a two-fold reduction in the training set size. 
If we were to train now for a fixed number of epochs $N_\mathrm{epochs}$, that is, until the network encountered each data point $N_\mathrm{epochs}$ times during training, we expect a better performance with a larger $N_\mathrm{train}$.
In this case, the network receives more update iterations to minimize the \ac{MSE} objective, hence the performance gain.
For a fairer comparison, we track both the training and the test loss during training after each update step.
Hence, the total number of iteration steps is to be made a constant, i.e.\ on a training set of twice the size we allow the network to train for half the epochs.
In this setting, each training run allows the \ac{NN} the same total amount of update steps.

\begin{figure}[h]
\centering \includegraphics[width=0.49\textwidth]{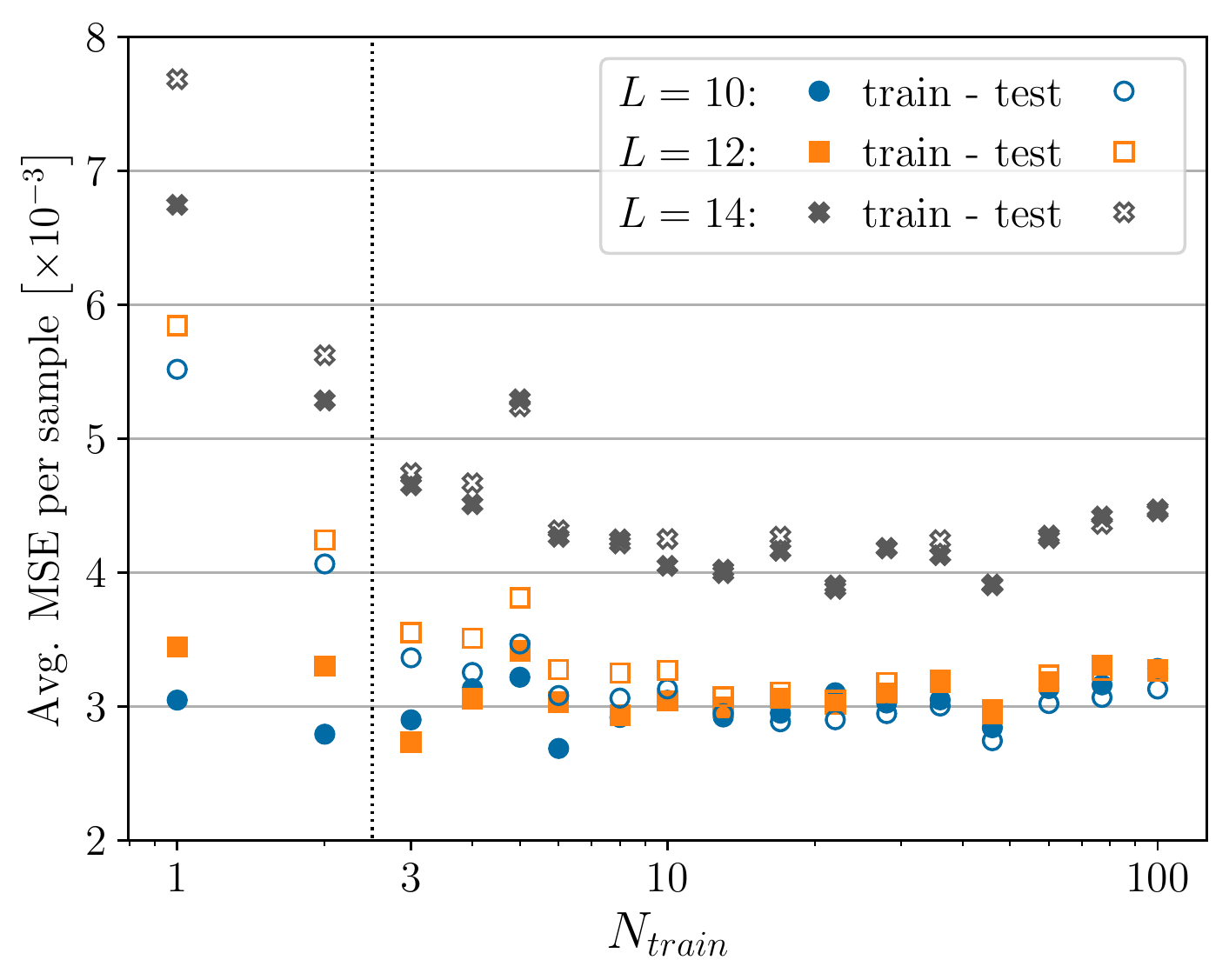}
\caption{
Dependence of the training and the test loss on the size of the corresponding training set.
$N_\mathrm{train}$ denotes how many realizations for each disorder parameter $h$ and chain length $L$ have been included in the training set.
Both losses are reported after convergence (around $1.350$ update steps).
We distinguish losses for different system size by color and the train from the test loss by different symbols, respectively.
\newline
We have averaged over ten independent training procedures.
There is no qualitative improvement for a training set with $N_\mathrm{train} \geq 3$ (vertical, dotted line).
Below this threshold, the network tends to overfit the available data, indicated by an increasing test error despite a decreased train error.
We have excluded data for $L=14$ from the training set, hence the increased losses for this system size.
\label{fig:N_train_dependence}
}
\end{figure}

In particular, this has resulted in very long training loops for a small $N_\mathrm{train}$ as we have trained for several hundreds of epochs.
Due to the mini-batching during training, we track the actual number of received update steps during training for various values of $N_\mathrm{train}$ and exclude the system size of $L=14$ from the training set.
We set a value of $N_\mathrm{epochs}=30$ for training on the largest data set size with $N_\mathrm{train} = 100$ and adjusted that value accordingly for smaller sizes.
In all considered cases, this leads to a convergence of the models and we extract the remaining average \ac{MSE} on both the training and the test set after convergence.
For each value of $N_\mathrm{train}$, we reinitialize and train the model ten times.
In all cases, when we decreased the training set, we have done so by always picking a random subset of the full training data set for each training reinitialization.
We show the two averaged losses in \cref{fig:N_train_dependence}.
This reveals that shrinking the training set down to $N_\mathrm{train} \approx 3$ (this corresponds to a total number of training points of around $180$) yields no qualitative increase of neither of the two losses after training.
This threshold is of the order or trainable parameters of the model (cf. \cref{fig:network_architecture}).
Below it, we observe a decreased training loss while the test loss is increased.
In this limit of scarce data, the model begins to overfit the training data at the expense of a larger loss on the test set.
This small number is encouraging for the model application to data that stems from an actual experiment as we have to repeat the same experiment only a handful of times for each point in the phase diagram we are interested in.
This highlights the feasability of our approach to actual data stemming from a quantum experiment.

\subsection{Further details on the phase diagrams}
\label{appendix:phase_diag}

In \cref{subsec:energy_dependency}, we highlight that our model is capable of dealing with various values for the energy density $\epsilon$.
Due to the choice of our architecture, $\epsilon$ is taken as an input feature for the subsequent fully-connected \ac{NN}.
We have experimented with various ways in presenting different values for $\epsilon$ to our model.
One initial alternative consists of various fully-connected \acp{NN} that are individually trained to predict the indicator values at a single $\epsilon$ each.
While this, at first, has appeared beneficial with respect to the validation loss, there are a few drawbacks of this approach.
The first one is the increased model complexity opposed to our scheme now.
Here, we only require one single \ac{NN} whereas the naive approach would require an \ac{NN} for every $\epsilon$ of interest.
Secondly, this approach limits the resolution of the prediction when in comes to obtaining the phase diagram in \cref{fig:phase_diag} as we require a data set for every $\epsilon$ of interest.
Our approach circumvents both issues by the introduction of $\epsilon$ as an intermediate feature.
This way, we can set up a much tighter grid for both $\epsilon$ as well as the disorder parameter $h$ and make predictions for each possible combination.
To this end, we sample $N=100$ new samples of disorder vectors $\Vec{h}$ for each $h$ and obtain the feature value by feeding it to the $\ac{RNN}$.
Then, we augment this value with every value of $\epsilon$ of interest and parse everything to the \ac{NN}.
Lastly, we average over $N$ and show this mean in dependence of $\epsilon$ and $h$ in the phase diagram.
Since we only require forward passes through our model, this procedure is highly efficient: the run time is proportional to the chain length $L$ and to the number of queried values for both $h$ and $\epsilon$ and in that sense optimal.

We have also experimented with analysing the model's predictions with a more quantitative measure such as the \ac{FSSA}~\cite{CCFSbound1986}.
This method is aimed at mitigating the finite-size effects in the data and to obtain quantitative estimates of the critical disorder parameter $h_c$ and the critical exponent of the transition $\nu$.
To this end, data from various chain lenghts is given to the \ac{FSSA} and fitted around the assumed value for $h_c$.
We have tried to query our model at chain lengths beyond those in the training set, i.e.\ $L>14$ but failed to reproduce previous approaches~\cite{LuLaflAlet2015} as we have not observed signs of the $\epsilon$-dependent mobility edge in the transition.
We attribute this observation to two different origins.
First, we observe that the approximation is of higher quality around the transition region (cf. \cref{fig:coeff_of_det}) and significantly so in the middle of the spectrum (at $\epsilon \approx 0.5$).
The latter might leave a bias in the data at either side of the spectrum which is observed in the phase diagram.
The second reason is due to our choice of the \ac{RNN} architecture as feature extractor.
In \cref{fig:typical_feature}, we have shown the typical feature vector produced by the \ac{RNN} after training.
One important aspect is that there exists a cross-over point that is independent of the chain length $L$ of the input data but whose position depends on the initialization of the network parameters.
This introduces a bias in the indicators since this cross-over is not apparent in the training data.
We have tried to average the output over multiple retrainings (and therefore feature vectors) and by increasing the number of features but failed to lift this bias.
However, we conjecture that with a more careful design of the \ac{RNN} architecture, this is possible.
In any case, the investigation of finding the right feature architecture is both interesting from a numerical and a theoretical perspective as it helps to shine some light on the nature of the \ac{MBL} transition.

\section*{Acronyms}

\begin{acronym}[POVM]\itemsep.5\baselineskip
\acro{AGF}{average gate fidelity}

\acro{BOG}{binned outcome generation}

\acro{CP}{completely positive}
\acro{CPT}{completely positive and trace preserving}
\acro{CS}{compressed sensing} 

\acro{DFE}{direct fidelity estimation} 
\acro{DM}{dark matter}

\acro{ETH}{eigenstate thermalization hypothesis}

\acro{FSSA}{finite-size scaling analysis}

\acro{GRU}{gated recurrent unit}
\acro{GST}{gate set tomography}

\acro{HOG}{heavy outcome generation}

\acro{LSTM}{long short-term memory}

\acro{MBL}{many-body localization}
\acro{ML}{machine learning}
\acro{MSE}{mean-squared-error}
\acro{MUBs}{mutually unbiased bases} 
\acro{MW}{micro wave}

\acro{NISQ}{noisy and intermediate scale quantum}
\acro{NN}{neural network}

\acro{POVM}{positive operator valued measure}
\acro{PVM}{projector-valued measure}

\acro{QAOA}{quantum approximate optimization algorithm}
\acro{QML}{quantum machine learning}
\acro{QPT}{quantum process tomography}

\acro{RNN}{recurrent neural network}

\acro{SFE}{shadow fidelity estimation}
\acro{SIC}{symmetric, informationally complete}
\acro{SPAM}{state preparation and measurement}

\acro{RB}{randomized benchmarking}
\acro{rf}{radio frequency}

\acro{TT}{tensor train}
\acro{TV}{total variation}

\acro{VQA}{variational quantum algorithm}

\acro{VQE}{variational quantum eigensolver}

\acro{XEB}{cross-entropy benchmarking}

\end{acronym}


\bibliographystyle{myapsrev4-1.bst}
\bibliography{references,mk}

\end{document}